\documentclass[lettersize,journal]{IEEEtran}
\usepackage{amsmath,amsfonts}
\usepackage{algorithmic}
\usepackage{algorithm}
\usepackage{array}
\usepackage[caption=false,font=normalsize,labelfont=sf,textfont=sf]{subfig}
\usepackage{textcomp}
\usepackage{stfloats}
\usepackage{url}
\usepackage{verbatim}
\usepackage{graphicx}
\usepackage{cite}

\usepackage{graphicx}
\usepackage{xcolor}
\usepackage{soul}
\usepackage{amssymb}
\usepackage[breaklinks=true]{hyperref}
\usepackage{breakcites}
\usepackage{makecell}
\usepackage{booktabs}
\begin{document}

\title{CLE-SH: Comprehensive Literal Explanation package for SHapley values by statistical validity}

\author{Kyungjin~Kim$^{1,2,3,4,*}$,
        Youngro~Lee$^{1,2,3,4,5,*}$,
        and Jongmo~Seo$^{1,2,3,4,6,\dagger}$%
\thanks{$^{1}$ Department of Electrical and Computer Engineering, Seoul National University, 1 Gwanak-ro, Gwanak-gu, Seoul, Republic of Korea.}%
\thanks{$^{2}$ Institute of Engineering Research, Seoul National University, Seoul, Republic of Korea.}%
\thanks{$^{3}$ Inter-University Semiconductor Research Center (ISRC), Seoul National University, Seoul, Republic of Korea.}%
\thanks{$^{4}$ Biomedical Research Institute, Seoul National University Hospital, Seoul, Republic of Korea.}%
\thanks{$^{5}$ Artificial Intelligence Institute, Seoul National University, Seoul, Republic of Korea.}%
\thanks{$^{6}$ Interdisciplinary Program of Medical Informatics, Seoul National University College of Medicine, Seoul, Republic of Korea.}%
\thanks{$^{*}$ These authors contributed equally to this work.}%
\thanks{$^{\dagger}$ Corresponding author: Jongmo Seo (email: callme@snu.ac.kr).}%
\thanks{This work was supported by the Institute of Information Communications Technology Planning Evaluation (IITP) Global Data X Leader HRD program grant funded by the Korean Government [Ministry of Science and ICT (MSIT)] under Grant IITP-2024-0044-1407.}%
}

\maketitle

\begin{abstract}
Recently, SHapley Additive exPlanations (SHAP) has been widely utilized in various research domains. This is particularly evident in application fields, where SHAP analysis serves as a crucial tool for identifying biomarkers and assisting in result validation. However, despite its frequent usage, SHAP is often not applied in a manner that maximizes its potential contributions. A review of recent papers employing SHAP reveals that many studies subjectively select a limited number of features as 'important' and analyze SHAP values by approximately observing plots without assessing statistical significance. Such superficial application may hinder meaningful contributions to the applied fields. To address this, we propose a library package designed to simplify the interpretation of SHAP values. By simply inputting the original data and SHAP values, our library provides: 1) the number of important features to analyze, 2) the pattern of each feature via univariate analysis, and 3) the interaction between features. All information is extracted based on its statistical significance and presented in simple, comprehensible sentences, enabling users of all levels to understand the interpretations. We hope this library fosters a comprehensive understanding of statistically valid SHAP results.
\end{abstract}

\begin{IEEEkeywords}
SHapley Additive exPlanations, Statistical Validity, Visualization
\end{IEEEkeywords}

\section{Introduction}
\label{sec:introduction}
\subsection{Background}

With the development of machine learning, the interpretation of models, specifically understanding the importance of each feature, has gained significant attention in biomedical areas. In these areas, datasets are often vulnerable to data bias, meaning high performance alone is not enough to validate the use of machine learning\cite{vokinger2021mitigating, CONEV2024108613, chen2024unmasking}. Interpretation that makes sense to experts can help reduce the potential risks of this bias. As machine learning utilizes non-linear and complex interactions between features that conventional regression analysis cannot handle, interpretation is often used to understand the mechanisms of target diseases\cite{hung2023developing, tso2022machine, zhang2020machine}. Additionally, interpretation is sometimes used as a primary goal to identify biomarkers among the vast numbers of genetic sequences or microbiomes\cite{wang2021characteristics, thomas2019metagenomic, lee2023machine, aryal2020machine}.

Interpretation of machine learning models built for tabular datasets can be done using two approaches: model-dependent and model-agnostic. The model-dependent approach includes indices such as permutation importance or the Gini-impurity of tree-ensemble algorithms like XGBoost and LGBM, which are popular these days. However, these measures provide only absolute importance, offering limited information compared to regression analysis. Therefore, the model-agnostic method, represented by SHapley Additive exPlanations (SHAP), is predominantly used for interpretation. SHAP analysis is an interpretable AI method based on game theory, elucidating how each feature affects predictions under various conditions. By providing explanations for each sample, it allows for a deeper understanding of feature importance, feature importance distribution by its feature value, and interaction analysis\cite{lundberg2017unified}.

Between April 1st, 2024, and April 7th, 2024, by searching for (SHAP[Title/Abstract]) AND (Machine Learning[Title/Abstract]) on the PubMed site, 38 papers that use machine learning and SHAP together were published. We analyze how papers in the biomedical fields utilize SHAP in Figure1 and Appendix Table 1. Among 33 accessible manuscripts, 84.8\% of the papers include SHAP summary plots, which summarize the SHAP value distribution for each feature by its rank \cite{aghababa2024exploration,fuse2024development,you2024development,fan2024predicting,he2024ultrasound,liu2024development,li2024insights,venturini2024predicting,yu2024exploring,abujaber2024machine,vimbi2024interpreting, wang2024quantifying, shinohara2024re, huang2024new,ma2024integrated,rodriguez2024machine,hernandez2024explainable,tong2024machine,guo2024application,wang2024exploring,meng2024application,li2024incremental,nagy2024predicting, rattsev2024incorporation, chen2024machine, lee2024essential, cheng2024early, chahine2024machine, xu2024interactive, pandey2024functional, gholi2024explainable, ciciora2024social, mallick2024game, cao2024explainable, liu2024improving, lun2024prediction, rajwa2024identification}.

\begin{figure*}
    \centering
    \includegraphics[width=0.9\textwidth]{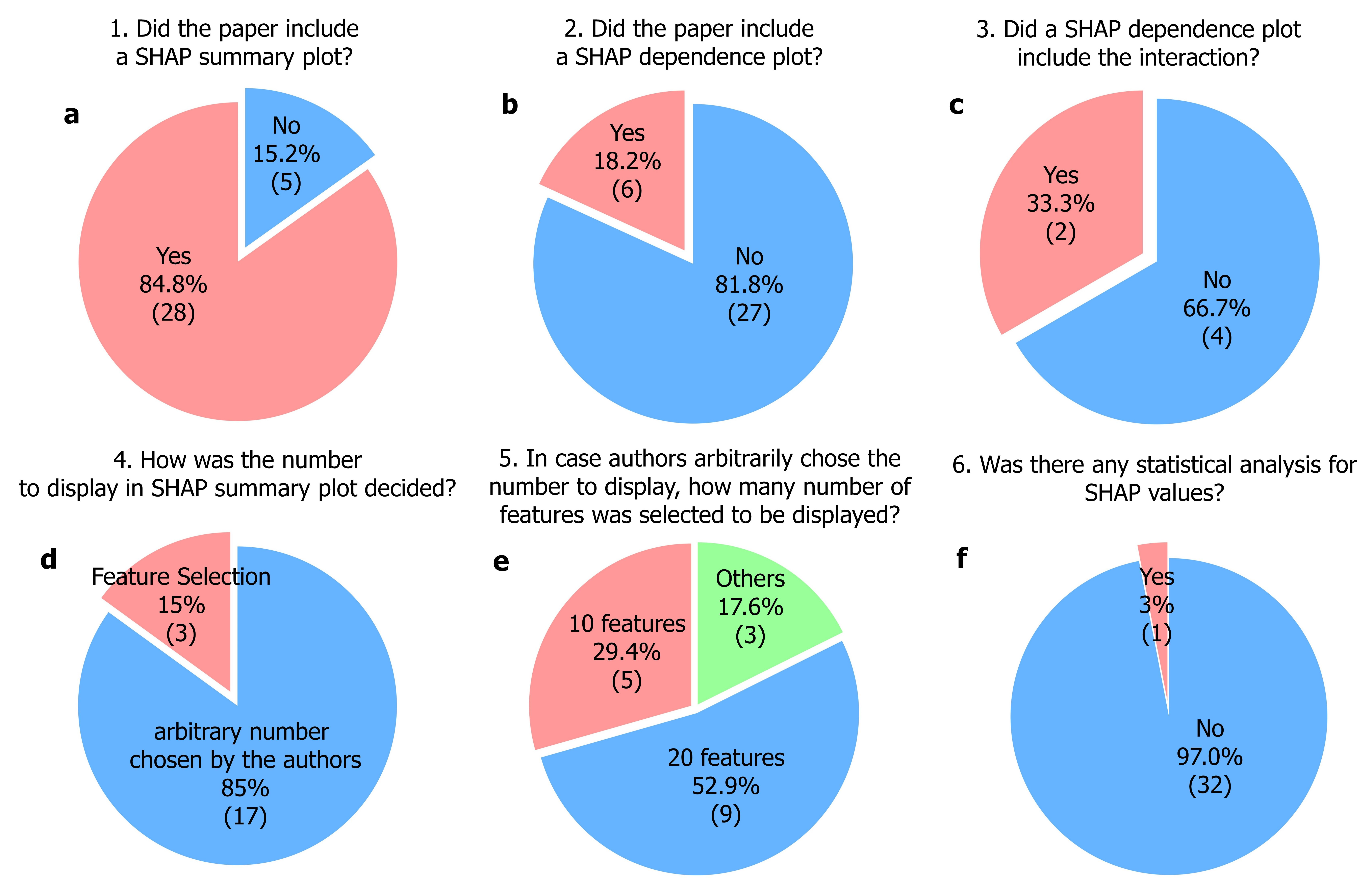}
    \caption{\textbf{Investigation on the SHAP analysis usage in recent papers in medical fields.} All figures were generated based on the search results available at the time of submission. One article (item 7 in Appendix Table 1) was retracted after submission. The figures and corresponding statistics were retained to preserve transparency of the initial dataset, but this retracted article had no influence on the study’s conclusions.}
\end{figure*}

\subsection{Problem Definition}
\label{subsec:problemDefinition}
We have identified three major issues with the current application of SHAP in state-of-the-art research.

\textbf{1) Lack of statistical validation}: Validation is critical in any field, especially in medicine. However, the current usage of SHAP in biomedical research lacks statistical validation and is often arbitrary. For example, Figure 1.d shows that among papers that did not display all features in the SHAP summary plot, only 15\% used feature selection methods. The majority chose arbitrary numbers like 10, 15, or 20 features without justification (Figure 1.e). Only one paper applied statistics to SHAP values, using them for Principal Component Analysis rather than to validate SHAP results \cite{rodriguez2024machine}. Without statistical tests, biomedical experts, particularly those unfamiliar with machine learning, may find it difficult to trust and utilize SHAP results.

\textbf{2) Complexity of interpretation}: While SHAP summary plots and other visualizations may be intuitive for data experts, they can be challenging for medical experts to interpret. The mixed colors in SHAP summary plots or SHAP dependence plots can obscure clear patterns, making it difficult to discern even highly ranked features.

\textbf{3) Absence of non-linear analysis}: SHAP is often used similarly to traditional feature importance methods, focusing only on ranking features. This neglects the unique capability of SHAP to provide interactive analysis through SHAP dependence plots. As shown in Figure 1.b, 81.8\% of papers did not include SHAP dependence plots. Even among the 18.2\% that did, most labeled scatter plots of single features as SHAP dependence plots.

These factors hinder the utility of SHAP analysis in providing interactive analysis and being referenced by medical experts who are not familiar with machine learning. However, there is no clear consensus or set of tools on how to extract statistically valid results from SHAP with comprehensive illustrations. Considering the different categories of feature variables and the varying patterns of SHAP values by feature, it is difficult and cumbersome for each researcher to do this individually.

\subsection{Related Works}
With the increasing popularity of SHAP analysis, there has been a growing number of studies leveraging SHAP values for deeper and more meaningful analyses. In this section, we summarize recent studies that address at least one of the issues outlined in Subsection~\ref{subsec:problemDefinition}.

\subsubsection{SHAP Values for Feature Selection}
\label{sec:4}

As SHAP values have gained attention, numerous attempts have been made to utilize them for feature selection. As shown in Figure 1, most approaches adapt traditional rank-based filtering methods to SHAP values. However, some studies propose entirely new methodologies for feature selection based on SHAP.

\begin{itemize}
    \item \textbf{BorutaSHAP}: Similar to the permutation importance approach, BorutaSHAP creates shadow features by randomly shuffling input features using TreeSHAP. It observes changes in performance across multiple iterations and calculates p-values to assess significance \cite{ekeany2020borutashap}.
    \item \textbf{Shapicant}: This method permutes the labels instead of features. It first calculates SHAP values for the actual labels and then computes null SHAP values for permuted labels. By comparing these, it calculates p-values to determine statistically significant differences \cite{calzolari2020shapicant}.
    \item \textbf{PowerSHAP}: PowerSHAP introduces noise into the dataset across multiple iterations and computes SHAP values. It evaluates the significance of SHAP values by comparing them to the added noise, identifying features with consistently high SHAP scores \cite{verhaeghe2022powershap}.
    \item \textbf{LLPowerSHAP}: Building on PowerSHAP, this method assumes that informative features will have significantly higher SHAP values compared to noise. It incorporates Logistic Loss SHAP and utilizes unseen data to identify informative features with minimal noise among the selected feature set \cite{madakkatel2024llpowershap}.
\end{itemize}

While these approaches suggest advanced feature selection methods with concrete algorithms, we propose a feature selection algorithm that also utilizes SHAP values. Our approach addresses the limitation of existing methods, which often require additional computational steps for feature selection. This requirement conflicts with our goal of generating a report file solely based on the input features and their SHAP values. In Section~\ref{sec:featureSelection}, we introduce a much lighter version of SHAP-based feature selection that eliminates the need for any supplementary experiments beyond the provided SHAP values.

\subsubsection{Analysis on SHAP Values}

There are numerous studies that perform additional analyses on SHAP values. Statistical results derived from linear and logistic regression was used to guide feature selection\cite{kraev2410shap}. The SHAPley EXplanation Randomization Test (SHAP-XRT) utilizes the Conditional Randomization Test (CRT) to assess the statistical significance of SHAP values in binary classification tasks, demonstrating how SHAP values can be employed to evaluate feature importance\cite{teneggi2022shap}.

In addition, several studies investigate the inherent instability of SHAP values. Similar to observations in other feature importance methods\cite{lee2024validity}, research has analyzed how SHAP values fluctuate due to random sampling\cite{yuan2022empirical}. Furthermore, other studies evaluate the validity of SHAP scores by examining their behavior within uncertainty regions, identifying the actual tight range of SHAP scores\cite{cifuentes2024distributional}.

In Section~\ref{sec:statistics}, we provide a list of statistics incorporated into our Python library. While each method represents well-known basic statistical techniques, to the best of the author's knowledge, no prior studies have applied such a wide range of statistical analyses to SHAP values.

\subsubsection{Simpler Explanation of SHAP Values}

Despite the widespread use of SHAP values, their interpretation often remains challenging for non-experts and can be subject to individual bias. To address this, several studies attempt to simplify SHAP explanations using large language models (LLMs). For example, one study uses pre-trained LLMs to analyze sample-specific risk indicators, providing enhanced interpretability \cite{zeng2024enhancing}. Another line of research employs LLMs to generate textual explanations for SHAP plots, making the visual outputs more accessible \cite{hsu2024decoding}.

While there have been attempts to interpret ML models in a textual context, there is still no consensus on which information to prioritize among the entire set of SHAP values. We believe that a deterministic step to extract the "important" information from SHAP values should precede the process of illustrating it in a textual manner.

% Although the depth of these studies significantly contributes to the progress in utilizing and understanding SHAP values, there are no comprehensive studies yet that address all the issues identified in \textbf{Section B}. In this study, we aim to address a wide spectrum of the mentioned problems by employing very simple statistical tests.

\subsection{Proposed Approach}

\begin{figure*}
    \centering
    \includegraphics[width=\textwidth]{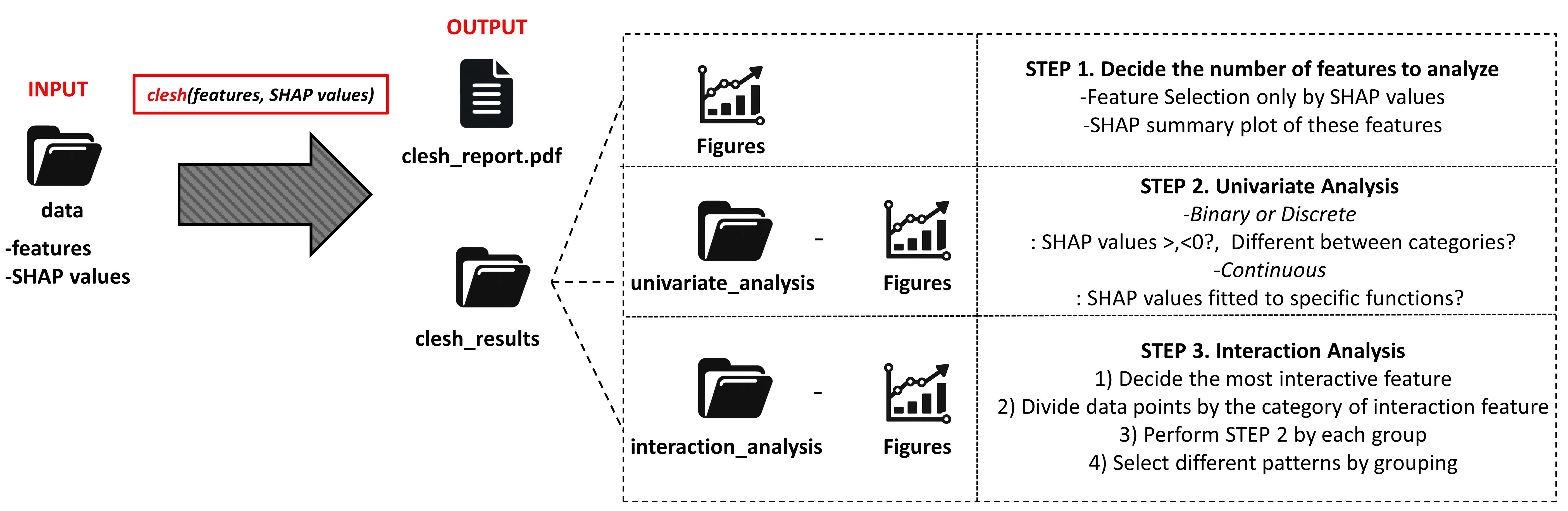}
    \caption{\textbf{Schematic of the library}}
\end{figure*}

The objective of this study is to develop a Python library, named CLE-SH, that automatically performs statistically validated SHAP analyses on biomedical tabular data and generates interpretable reports for both experts and non-experts.
Rather than serving as a final step that replaces all analyses, CLE-SH is intended to be used in the early exploratory stage, helping researchers design a clear analytical direction from complex, information-rich datasets.
To achieve this objective, we designed the library to integrate statistical validation, feature-type classification, univariate and interaction analyses, and report generation within a unified automated pipeline.

\textbf{1) Statistical tests to analyze SHAP values}: In Sections~\ref{sec:featureSelection},\ref{sec:featureType},\ref{sec:univariateAnalysis},\ref{sec:interactionAnalysis}, we introduce a procedure to analyze SHAP values with statistical significance for any type of tabular dataset. First, we define the number of important features to focus on, which involves feature selection by SHAP values (Section~\ref{sec:featureSelection}). Next, we determine the data type of each important feature, whether it is binary, discrete, or continuous (Section~\ref{sec:featureType}). Considering the type of each feature, we suggest univariate analysis to analyze each feature individually (Section~\ref{sec:univariateAnalysis}). Finally, we design an interactive analysis to filter out statistically significant results from SHAP dependence plots (Section~\ref{sec:interactionAnalysis}).

\textbf{2) Summary report generation}: We summarize the statistically significant results into a report intended to be comprehensive for non-experts in data analysis. To achieve this, instead of only displaying plots, we also generate literal sentences that explain the significant patterns within the plots (Appendix 6)

\section{Methods}
\label{sec:methods}
\subsection{Datasets}
To validate the generalizability and reproducibility, we test our package using examples from five different datasets from various biomedical areas. The data sources and label definitions for each dataset are described in Appendix 1, while detailed information on the number of samples, features, feature types, and obtainable performance is summarized in Table 1, demonstrating the diversity of the datasets \cite{albert2024_metabolic, patel2015reliable, https://doi.org/10.24432/c5z89r, antal2014diabetic, lee2023machine}. We collect SHAP values by test features in each 5-fold cross-validation, ensuring that every data sample was included in the results. The XGBoost algorithm, a widely used tree-based ensemble method popular in biomedical research, is used for the machine learning model (Appendix 2). We use the area under the ROC curve (AUC) to calculate the performance for each dataset. Details of our testing environment are provided in Appendix 3.

In this study, our purpose is to test the results of our library, rather than to interpret them in terms of biological or clinical knowledge. Therefore, we do not include any analysis of the results from that perspective, and some features are referred to by their feature numbers rather than names. When users actually use this package, if the input data file includes the names of the features, the results will be generated based on those names.

\begin{table}[h!]
    \centering
    \caption{\textbf{Data Characteristics}.  MS*=Metabolic Syndrome, BC*=Breast Cancer, HF*=Heart Failure, DR*=Diabetic Retinopathy, IBD*=Inflammatory Bowel Disease}
    \small
    \begin{tabular}{|p{2cm}|c|c|c|c|c|}
        \hline
        \textbf{Label Name}  & \textbf{MS*} & \textbf{BC*} & \textbf{HF*} & \textbf{DR*} & \textbf{IBD*}  \\
        \hline
        \textbf{\# of samples} & 2009 & 198 & 299 & 1151 & 1569   \\
        \hline
        \textbf{\# of positive} & 822  & 47  & 96  & 611  & 702  \\
        \hline
        \textbf{Average AUC} & 0.96  & 0.72  & 0.90 & 0.77 & 0.99   \\
        \hline
        \textbf{\# of features} & 17 & 31 & 12 & 19 & 283   \\
        \hline
        \textbf{Binary} & 7 & 0 & 5 & 3 &  0  \\
        \hline
        \textbf{Discrete} & 1 & 0 & 1 & 0 &   0 \\
        \hline
        \textbf{Continuous} & 9 & 31 & 6 & 16 & 283  \\
        \hline
    \end{tabular}
\end{table}

\subsection{Statistics}
\label{sec:statistics}

This study primarily involves grouping SHAP values based on feature values and analyzing these grouped SHAP values or comparing them across groups. The statistical experiments utilized in this study are as follows:

\subsubsection{Sign of SHAP Values}

This experiment examines whether the SHAP values of a specific group significantly lean toward a particular sign (positive or negative), to decide whether a feature consistently contributes positively or negatively to the model prediction across certain value ranges.

\textit{One-sample t-test:} This test checks whether the mean of SHAP values in a group is significantly different from zero. It assumes that the SHAP values are normally distributed. If the mean is significantly positive or negative, it provides evidence of the overall directionality of the SHAP values in that group.

\textit{Wilcoxon signed-rank test:} As a non-parametric alternative, this test is used when the normality assumption is violated. It evaluates whether the median of SHAP values differs from zero, making it robust to outliers and skewed data distributions.

\subsubsection{SHAP Values Between Two Groups}

This experiment assesses whether the differences in SHAP values between two groups are statistically significant. 

\textit{Two-sample t-test:} This parametric test compares the means of SHAP values between two independent groups. It assumes normality and equal variances across the groups. A significant result indicates that the mean SHAP value differs between the groups, suggesting that the feature's contribution to the model prediction varies based on the group.

\textit{Mann-Whitney U test:} When normality cannot be assumed, this non-parametric test is employed to compare the distributions of SHAP values between two groups. It is particularly useful when the data are ordinal or contain outliers, as it assesses whether one group tends to have higher or lower SHAP values than the other.

These tests are often applied to compare groups defined by binary feature values, such as comparing SHAP values for "Male" vs. "Female" in gender-related features.

\subsubsection{SHAP Values Between Multiple Groups}

This experiment evaluates whether there are significant differences in SHAP values across more than two groups.

\textit{One-way ANOVA:} This parametric test compares the means of SHAP values across multiple groups. It assumes that the data are normally distributed and have equal variances across groups. A significant result indicates that at least one group mean differs from the others.

\textit{Kruskal-Wallis test:} When the assumptions of ANOVA are not met, this non-parametric alternative is used to compare the medians of SHAP values across groups. It evaluates whether the distributions of SHAP values differ significantly among the groups.

\textit{Post-hoc tests:} If ANOVA or Kruskal-Wallis tests reveal significant differences, post-hoc tests (e.g., Tukey’s HSD for parametric data or Dunn’s test for non-parametric data) are conducted to identify which specific groups differ from each other. In this study, we utilized Tukey's HSD as a post-hoc test.

\subsubsection{Regression Analysis}

This experiment focuses on fitting SHAP values for continuous feature values to specific functions to evaluate the relationship between feature values and their contributions to model predictions.

\textit{Procedure:} Feature values are set as the input variable (\textit{x}), and SHAP values are the output (\textit{y}). Each predefined function (e.g., linear, quadratic, exponential) is fitted to the data. The statistical significance of each function is evaluated based on the p-value of the coefficient of \textit{x}, denoted as \textit{a}.

\textit{Selection of the best function:} If multiple functions are found to be statistically significant, the Root Mean Square Error (RMSE) is used to identify the function that best fits the data. This approach ensures that the selected function accurately represents the relationship between the feature and its SHAP values.

\textit{Applications:} This analysis is valuable for identifying whether the impact of a feature is linear, non-linear, or follows a more complex pattern. For example, it can reveal whether an increase in a feature's value consistently increases or decreases its contribution to the model, or if there are threshold effects.

\subsection{How to use the library}
The description and installation guide for the CLE-SH package is accessible at: \underline{https://github.com/kyungjini/cle-sh}. After installation, importing the package with \texttt{import clesh} will enable the analysis by executing \texttt{clesh(features, SHAP values, label name)} as in Figure2. Here, \textit{features} indicates the dataframe containing the feature dataset used to train/test the machine learning model and calculate SHAP values. \textit{SHAP values} are SHAP values which match to \textit{features}. \textit{label name} is required to generate more comprehensive literal sentences.

\section{Number of important features}
\label{sec:featureSelection}

\begin{figure*}
    \centering
    \includegraphics[width=\textwidth]{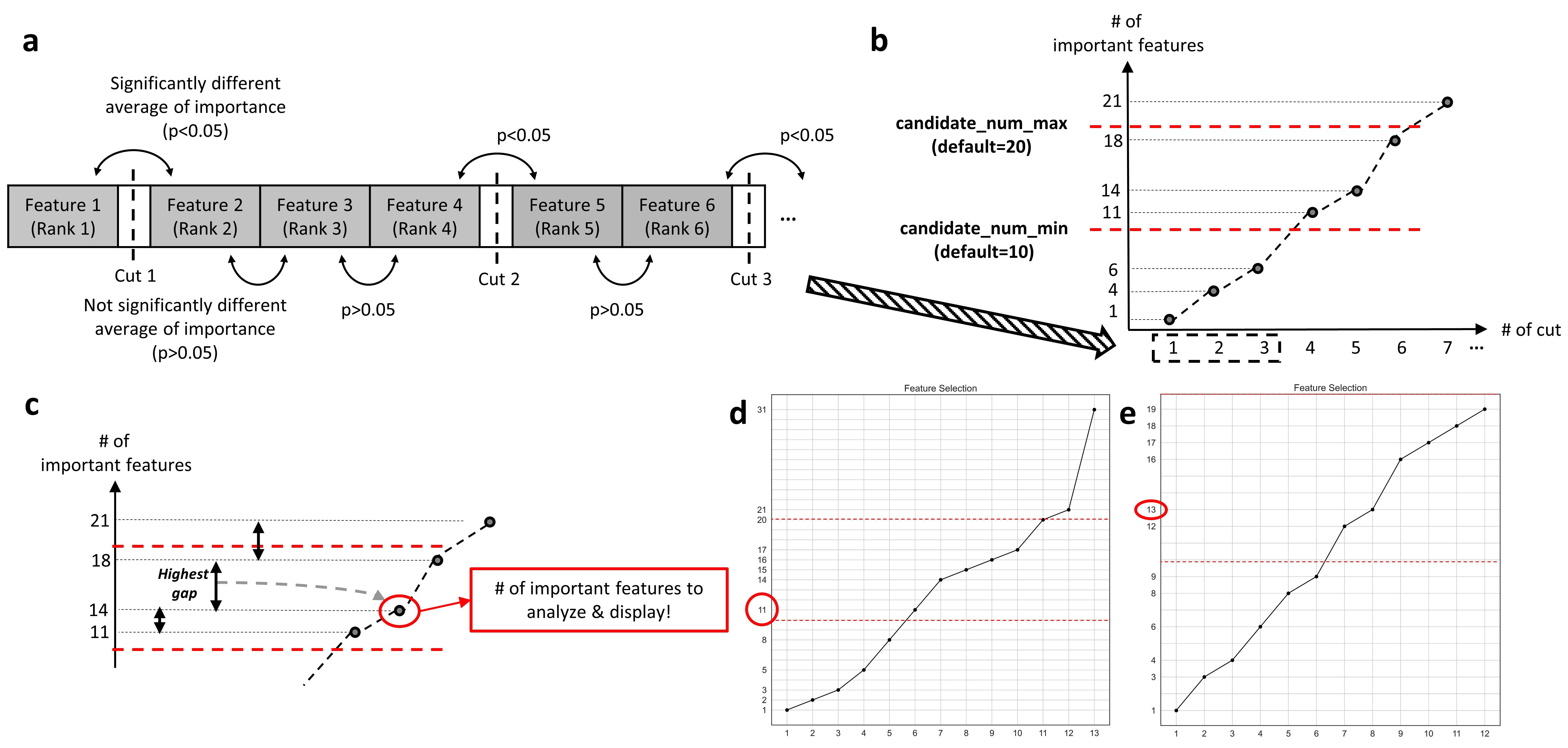}
    \caption{\textbf{Graphical explanation of determining the number of important features using SHAP with examples}. a-c. Visual explanation of the feature selection process. d. Feature selection plot derived from the BC dataset and e) from the DR dataset.}
\end{figure*}

Displaying and analyzing every feature might be inefficient as not all features are important and it can harm the effective delivery. As shown in Figures 1.d and 1.e, the number of features displayed in SHAP summary plots is usually decided to be 10 or 20 without validation, such as feature selection. Although justifying the number of important features is crucial, it is very cumbersome for analysts to perform feature selection every time. This inconvenience arises particularly because justifying with the best model algorithm might increase the computational burden. As a result, even when feature selection is used to validate the number of important features to analyze, the algorithm is often based on much simpler and smaller models that are not related to SHAP values or the final predictive model used to collect SHAP values. To address this issue, we use a feature selection method oriented around SHAP values \cite{lee2023suggestion}. By automating this process, the justification for the number of important features is conveniently provided. 

As statistical significance is a key factor in this study, we compare the average absolute SHAP values between features. As shown in Figure3.a, we calculate the statistical significance of difference between adjacently ranked features. When both groups of SHAP values satisfy the normal distribution test by the Shapiro Wilk test, a paired t-test is used (parametric); otherwise, the Wilcoxon rank sum test is used (non-parametric). By cutting the group of features by where statistical significance occurs, we determine the number of important features according to the number of cuts, as shown in Figure3.b. The minimal number of important features (\textit{candidate\_num\_min}, default=10) and the maximal number of important features (\textit{candidate\_num\_max}, default=20) are set by hyperparameters, as having too many important features can be ineffective and too few can be less informative. Between \textit{candidate\_num\_min} and \textit{candidate\_num\_max}, we select the number that shows the highest difference from the next cut, as shown in Figure3.c. Statistical significance is determined p-value<0.05, which 0.05 can be changed by adjusting the hyperparameter \textit{p\_feature\_selection}. In the \textbf{clesh\_result} folder (see Figure2), the Figureshowing the number of important features according to the number of cuts, as in Figures 3.d and 3.e, will be saved.

\section{Identification of feature data type}
\label{sec:featureType}
Defining the type of features is necessary to design an appropriate statistical test pipeline. In this paper, we define three different types of features: binary, discrete, and continuous. By calculating the unique values of each feature, the feature type can be easily determined. If the unique feature values are only two, the feature is classified as binary. If the list of unique feature values exceeds a defined threshold (hyperparameter \textit{cont\_bound}), the feature is classified as continuous; otherwise, it is classified as discrete. We set the default for \textit{cont\_bound} to 10. In Table 1, the number of features for each data type is listed.

\section{Univariate analysis}
\label{sec:univariateAnalysis}
The overall process of univariate analysis is illustrated in Figure 4, highlighting the statistical tests and interpretation flow by feature type.

\begin{figure}
    \centering
    \includegraphics[width=\columnwidth]{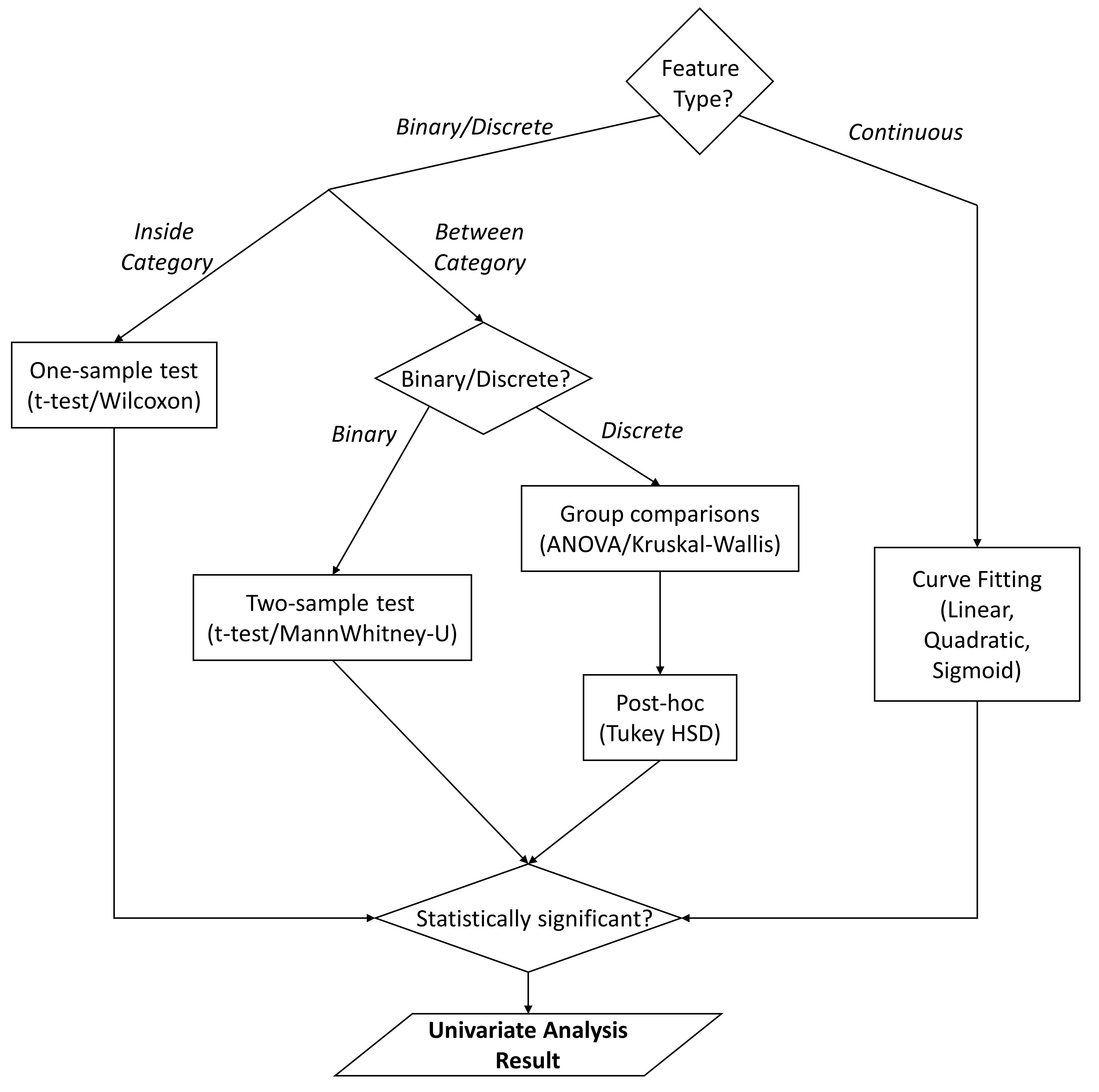}
    \caption{\textbf{Flowchart of the univariate SHAP value analysis pipeline}. This flow outlines the analysis procedure by feature type (binary, discrete, and continuous), showing how SHAP values are statistically processed and interpreted using hypothesis tests and curve fitting.}
\end{figure}

Based on the type of each feature by Section 4, we plot figures which can generally describe the distribution of SHAP values, while analyzing the univariate impact with statistical significance. This analysis is done only for important features selected in Section 3, but can be expanded by setting the hyperparameter \textit{manual\_num}. The standard for statistical significance in this section can be controlled by the hyperparameter \textit{p\_univariate}. In Table 2, we summarize the result of univariate analysis for five datasets. More examples like as Figure 5 are illustrated in Appendix 4. Any figures drawn in this section are saved in the folder,  \textbf{clesh\_results/univariate\_analysis} as in Figure 2.

\begin{table}[h!]
    \centering
    \caption{\textbf{Univariate analysis in each dataset}. Linear, Quadratic, and Sigmoid in the first column imply the number of features that fit best to each function with statistical significance.}
    \small
    \begin{tabular}{p{3.2cm} c c c c c}
        \toprule
        \textbf{Label Name}  & \textbf{MS*} & \textbf{BC*} & \textbf{HF*} & \textbf{DR*} & \textbf{IBD*} \\
        \midrule
        \textbf{\# of important features} & 17  & 11  & 12  & 13 & 12   \\
        \cmidrule{1-6}
        \textbf{Binary} & 7 & 0 & 5 & 0 & 0   \\
        \textbf{Discrete} & 1 & 0 & 1 & 0 & 0   \\
        \cmidrule{1-6}
        \textbf{Continuous} & 9 & 11 & 6 & 13  & 12   \\
        \cmidrule{1-6}
        Linear &  0 & 3 & 0 & 0 &  0  \\
        Quadratic & 2 & 2 & 2 & 3 & 0   \\
        Sigmoid & 7 & 6 & 4 & 10 &  12  \\
        \bottomrule
    \end{tabular}
\end{table}

\subsection{Binary and Discrete Types}
\label{subsec:uni-disc}
% SHAP dependence plots without interaction analysis, as shown in previous papers (Figure2.e), can illustrate the distribution of SHAP values in each category. However, since it is a scatter plot, understanding the distribution by the density of dots is challenging, and identifying the average point is difficult. To minimize subjective interpretation, 
We provide box plots of SHAP values in each category, as shown in Figures 5.a and 5.b.
% Before analyzing the size of SHAP values, the sign should be checked.

\textbf{Inside category}: To understand an overview of how each category influences the label, SHAP values of each category are tested to determine whether the average is significantly higher or lower than zero. For each group, normality is tested using the Shapiro-Wilk test; if normal, an one sample t-test is applied, and if not, the Wilcoxon signed rank-sum test is used. 

\textbf{Between categories}: The distribution of importance in each categories is compared to understand which category shows different pattern from another. For binary features, two sample t-test is applied when both groups are normal by Shapiro Wilk test, and if not, Mann-Whitney U test is applied. For discrete features, one-way ANOVA (parametric) or the Kruskal-Wallis test (non-parametric) is used to identify any statistically significant differences between categories. If a significant difference is found, Tukey's HSD is used to identify which categories differ significantly ($\textit{p}<0.05$). The result of Tukey's HSD is represented as in Figure 5.c. 

\subsection{Continuous Types}
\label{subsec:uni-cont}

SHAP values of continuous features are difficult to interpret using conventional statistics, as the relationship can be very complex to interpret or feature values can be too sparse. To balance the amount of information and utility, we first provide scatter plots (SHAP dependence plots without interaction analysis). Then, we apply three different functions for interpretation: linear (Eq. 1), quadratic (Eq. 2), and sigmoid (Eq. 3). By setting feature values as the input variable and SHAP values as the output, each function is fitted and the function is selected as in Section~\ref{sec:statistics} 2) Regression Analysis. The selected function is displayed in the scatter plot. As shown in Table 2, the function type that fits the most is the sigmoid function, followed by quadratic and linear. This indicates that the relationship between feature values and SHAP values is usually non-linear, underscoring the need to analyze more than just the SHAP summary plot. An example of the linear function is illustrated in Figure 5.d, the quadratic function in Figure 5.e, and the sigmoid functions in Figures 5.f and 5.g.

\[
f(x) = ax + b \quad \text{(Eq. 1)}
\]

\[
f(x) = ax^2 + bx + c \quad  \text{(Eq. 2)}
\]

\[
f(x) = \frac{L}{1 + \exp(-a(x - x_0))} + b \quad  \text{(Eq. 3)}
\]

\begin{figure*}
    \centering
    \includegraphics[width=0.65\textwidth]{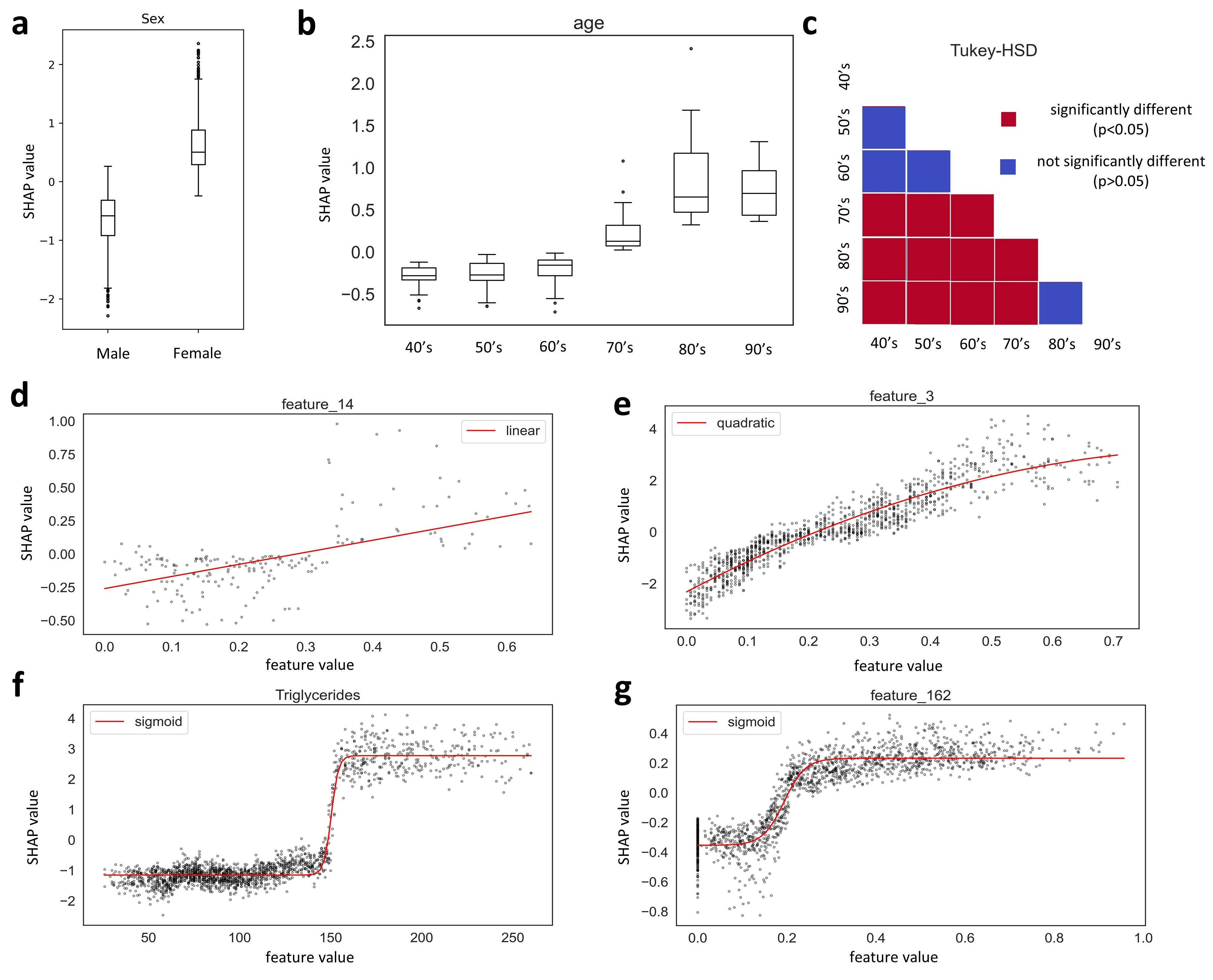}
    \caption{\textbf{Examples in univariate analysis}. a. Binary type (Dataset: MS), b. Discrete type (Dataset: HF), c. Result of Tukey-HSD test (Feature from Figure4b), d. Continuous type fitted to a linear function (Dataset: BC), e. Continuous type fitted to a quadratic function (Dataset: DR), f. Continuous type fitted to a sigmoid function (Dataset: MS), g. Continuous type fitted to a sigmoid function (Dataset: IBD). Other examples can be found in Appendix 4}
\end{figure*}

\section{Interaction analysis}
\label{sec:interactionAnalysis}
\begin{figure}
    \centering
    \includegraphics[width=\columnwidth]{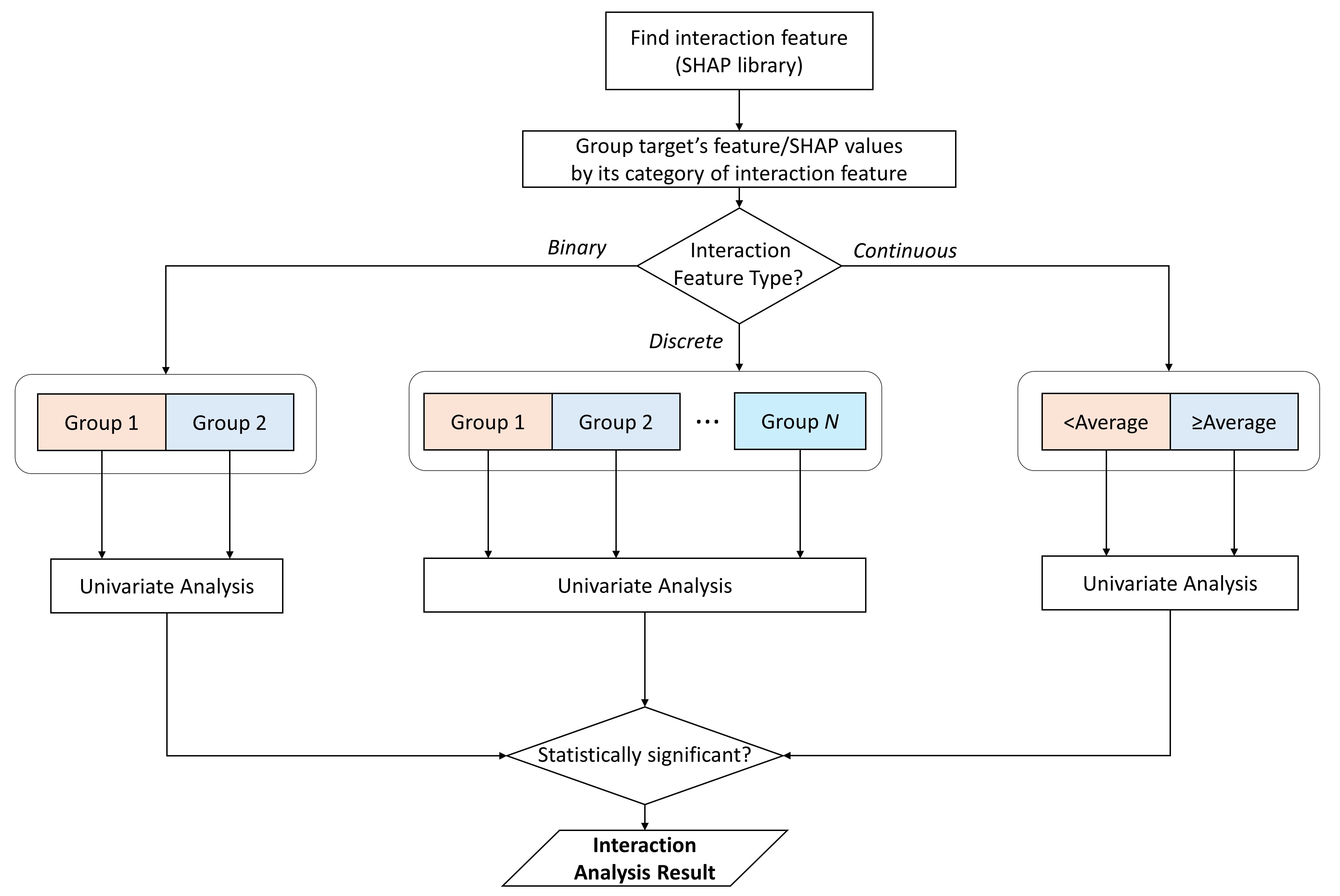}
    \caption{\textbf{Flowchart of the interaction analysis pipeline}. The diagram illustrates how combinations of target and interaction feature types (binary, discrete, continuous) are handled, from feature pairing and grouping to significance testing and final reporting.}
\end{figure}

The overall process of interaction analysis is illustrated in Figure 6, showing how our pipeline systematically handles various combinations of target and interaction feature types. The analysis is performed for important features identified in \mbox{Section~\ref{sec:featureSelection}}, referred to as target features. For each target, we define an interaction feature using the built-in SHAP utility function \mbox{\textit{shap.utils.approximate interactions().}}

Interaction analysis is performed for important features in Section~\ref{sec:featureSelection}, namely, target features. To define the interaction feature of each feature, we use the built-in function of SHAP library, \textit{shap.utils.approximate\_interactions()}. 

Although interaction analysis is a unique strength of SHAP compared to other feature importance methods, defining interactions can be challenging due to the numerous combinations of data types and relationships. We believe that this complexity, along with the inconvenience of manually examining each SHAP dependence plot, limits the potential of SHAP's interaction analysis (Figures 1.b and 1.c). Rather than overwhelming users with information, we aim to simplify the pipeline and recommend specific interactions to focus on by illustrating the statistical significance of certain patterns. We plot the SHAP values of the target feature separately based on the interaction feature's value as in Figure 7. Any figures drawn in this section are saved in the folder,  \textbf{clesh\_results/interaction\_analysis}. Additionally, we perform a statistical test to determine whether the value of the interaction feature has a significant impact on the SHAP values of the target feature. The standard for statistical significance in this section can be controlled by the hyperparameter \textit{p\_interaction}.

\begin{figure*}
    \centering
    \includegraphics[width=0.8\textwidth]{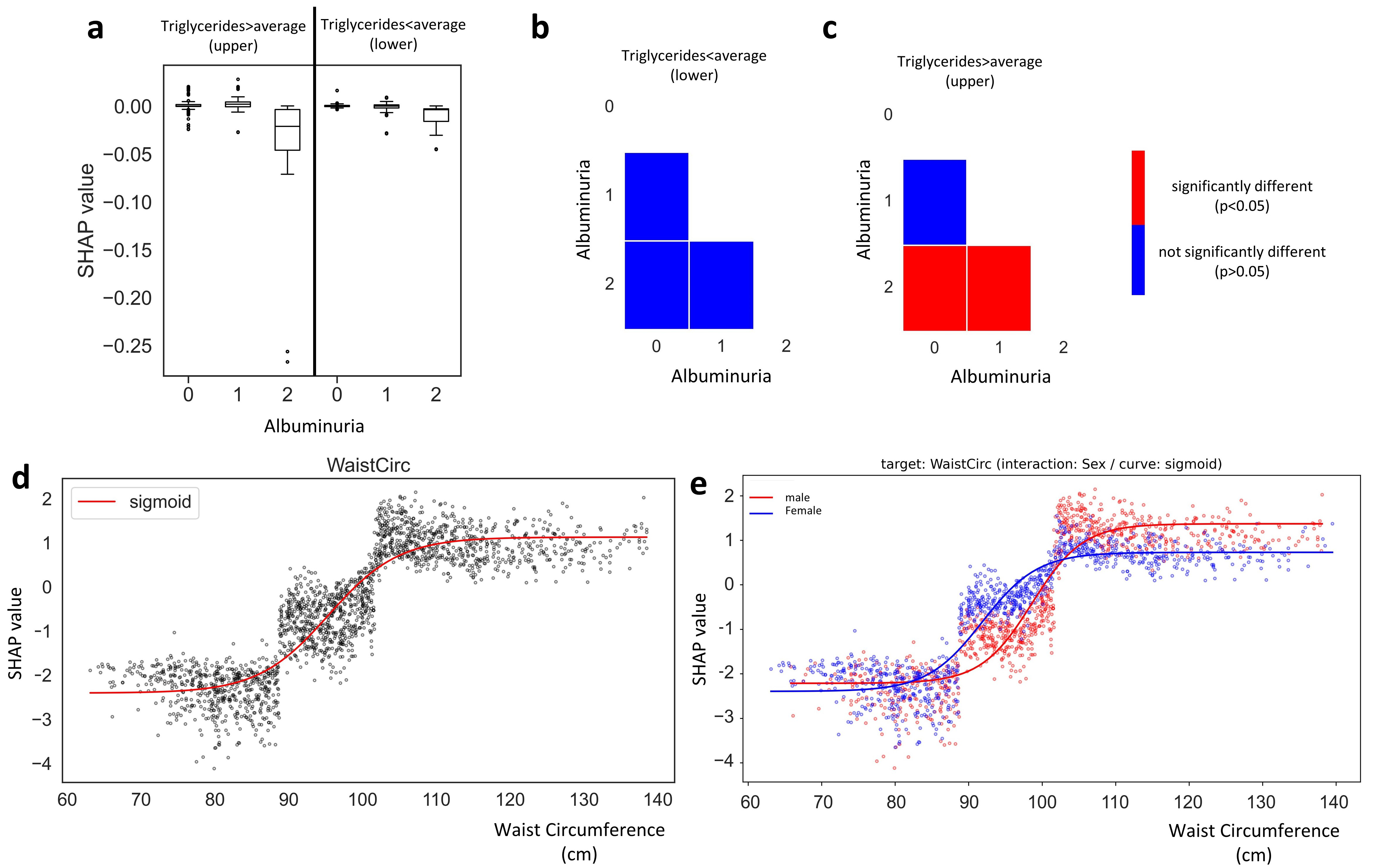}
    \caption{\textbf{Examples in interaction analysis}. a. Box plot of 'Albuminuria' with the interaction feature 'Triglycerides' (Dataset: MS). b. Tukey's HSD result for samples where Triglycerides is lower than the average.
    c. Tukey's HSD result for samples where Triglycerides is higher than the average. d. Scatter plot of Waist Circumference with a sigmoid function fitted by univariate analysis (Dataset: MS). e. Scatter plot of Waist Circumference with the interaction feature 'sex'. Other examples can be found in Appendix 5}
\end{figure*}

\subsection{Target: Binary/Discrete, Interaction: Binary/Discrete.}
\label{subsec:disc-disc}
In each category of the interaction feature, the SHAP values of the target feature between different categories of the target feature are compared. The statistical test follows the method described in Section~\ref{subsec:uni-disc}. When a target feature is binary, a two-sample t-test (parametric) or Mann-Whitney U test (non-parametric) is applied. When a target feature is discrete with multiple categories, a one-way ANOVA (parametric) or Kruskal-Wallis test (non-parametric) is used, followed by Tukey’s HSD post-hoc analysis to identify statistically significant differences between categories.

\subsection{Target: Binary/Discrete, Interaction: Continuous.}
\label{subsec:disc-cont}
When the interaction feature is continuous, analyzing by its range can be difficult to interpret. To simplify, we group interaction values into two categories: those smaller than the average and those larger than the average. This allows us to treat the interaction feature as a binary feature and apply the analysis described in Subsection~\ref{subsec:disc-disc}.

In Figure 7.a, an example of box plots in interaction analysis is illustrated (dataset: MS). The target feature 'Albuminuria' shows a negative slope in its SHAP values, regardless of the interaction feature's value (Triglycerides). However, in Figures 7.b and 7.c, we observe that the SHAP values, where 'Albuminuria's feature value is 2, show a statistically significant difference from other SHAP values only when Triglycerides is above its average. This indicates that the impact of Albuminuria = 2 is more significant when triglycerides are above the average.

\subsection{Target: Continuous, Interaction: Binary/Discrete.}
\label{subsec:cont-disc}

SHAP values of the target feature are grouped by the category of the interaction feature. Each group of SHAP values is then regressed by the target feature values using the function type selected in Subsection~\ref{subsec:uni-cont}. Functions that are fitted with statistical significance will be displayed in the scatter plots as shown in Figure 7.d. If only one group (one category of the interaction feature) shows statistical significance for the fitting, there will be no comparison between categories of the interaction feature. If more than one group shows statistical significance for the fitting, a comparison is made to determine whether the pattern depends on the category of the interaction feature.

However, comparing two different function equations is complicated because the difference depends on which range of target feature values is compared. And analyzing the difference by its location complicates the interpretation and deviates from our goal of comprehensiveness. Therefore, we only analyze the overall margin between the fitted functions of each group. For every target feature value, we compare the SHAP value regressed by each group using a paired t-test (parametric) or Wilcoxon rank-sum test (non-parametric).

In Figure 7.d, although the scatter plot is fitted into sigmoid function, it is not sufficient to explain the three stages of distribution. Figure 7.e shows that SHAP values of waist circumference depend on sex, and the scatter plot in Figure 7.d represents the mix of two different sigmoid functions ($\textit{p}<0.05$).

\subsection{Target: Continuous, Interaction: Continuous.}
\label{subsec:cont-cont}

As in Subsection~\ref{subsec:disc-cont}, samples are divided into two groups based on whether the interaction value is below or above its average. The analysis follows the same procedure as in Subsection~\ref{subsec:cont-disc}, treating the interaction feature as binary.

\section{Discussion}
This study aims to support researchers who seek to use SHAP analysis in a statistically valid and reliable manner by generating summary reports based on statistical significance. 
CLE-SH is designed to guide users toward more rigorous and reproducible usage of SHAP at the early and mid stage of analysis. 
To achieve this objective, we implemented a minimal yet essential pipeline that ensures statistical validity throughout the interpretability process. 
Section~\ref{sec:featureSelection} describes the feature selection procedure, which identifies how many features should be primarily examined using SHAP values alone, without additional experiments. 
Section~\ref{sec:featureType} classifies each feature according to its variable type, an essential prerequisite for proper analysis. 
Section~\ref{sec:univariateAnalysis} conducts univariate analyses for individual features, 
and Section~\ref{sec:interactionAnalysis} examines whether statistically significant interactions exist with the most interactive feature. 
The overall structure and example template of the automatically generated report are provided in Appendix~6, and complete PDF reports for each dataset are included in the source code repository.
By integrating these modules into a unified workflow, CLE-SH bridges model explainability and statistical rigor. 
It enables users to obtain interpretable and statistically reliable explanations from SHAP values without manual configuration, thereby making SHAP analysis both rigorous and accessible.

The proposed library is intended to function primarily at the exploratory stage of data analysis rather than at the stage of final statistical confirmation.
Because the appropriateness of a specific analytical approach depends on the characteristics of each dataset, it is neither practical nor desirable to predetermine a fixed analysis pipeline.
Instead, CLE-SH aims to assist researchers in the early phase of investigation by summarizing statistically significant SHAP-based patterns, thereby helping them identify promising analytical directions.
For this reason, the current version focuses on fundamental and essential functionalities, which naturally come with certain limitations but also provide many possibilities for future development.

As discussed in the main text, once the analysis moves into the interaction domain, it becomes inherently more complex and highly dependent on the specific data context.
For continuous variables in particular, there may exist patterns beyond the functional forms considered in this study, and the presence or absence of interaction effects can vary depending on how variable boundaries are defined.
In the current version, we analyzed only the single most prominent interaction for each dataset. However, interactions naturally occur at multiple levels and can be explored from various perspectives.
For instance, Bayesian approaches could be adopted to model hierarchical dependencies among features, or large language models could be leveraged to generate more generalized statements about feature interactions.

As this library is intended to provide a summary report that can be used in the early exploratory and feature-screening stages of analysis, its applicability to large-scale medical data is an important consideration.
Although CLE-SH has not yet been directly tested on very large clinical datasets, the statistical procedures implemented are relatively lightweight, and no computational issues are expected.
However, since p-values are inherently dependent on sample size, very large datasets may yield inflated statistical significance even for negligible effect sizes.
This suggests that further refinement could be made in selecting appropriate statistical techniques or by incorporating additional warning functions that alert users when significance might be driven primarily by sample size rather than true effect magnitude.

It is also worth noting that SHAP values can vary depending on the underlying predictive model. In this study, XGBoost was selected as the representative model for SHAP computation. Although CLE-SH can process SHAP values derived from any algorithm, tree ensemble models tend to align more naturally with the library’s statistical framework due to their discrete partitioning of feature spaces and the deterministic nature of TreeSHAP explanations \cite{lundberg2019explainable, lee4744730biomarker}. In contrast, neural network based models, such as multilayer perceptrons, generally require approximation-based SHAP methods, which are computationally more demanding and can be more sensitive to model parameters or initialization \cite{slack2020fooling, adebayo2018sanity}. Addressing these challenges and extending CLE-SH to support a wider range of model architectures will be an important direction for future work.

However, as the functionality of the library expands, it will gradually diverge from its original purpose as a lightweight screening tool for early-stage observation.
Therefore, maintaining usability while incorporating more advanced analytical features will be a key challenge.
Future development should focus on improving the user interface and accessibility—such as implementing a web-based platform and integrating large language models—to make the library more comprehensive and intuitive for diverse users.


% \bibliographystyle{ieeetr}
% \bibliography{references}

\begin{thebibliography}{10}

\bibitem{vokinger2021mitigating}
K.~N. Vokinger, S.~Feuerriegel, and A.~S. Kesselheim, ``Mitigating bias in machine learning for medicine,'' {\em Communications medicine}, vol.~1, no.~1, p.~25, 2021.

\bibitem{CONEV2024108613}
A.~Conev, R.~Fasoulis, S.~Hall-Swan, R.~Ferreira, and L.~E. Kavraki, ``Hlaequity: Examining biases in pan-allele peptide-hla binding predictors,'' {\em iScience}, vol.~27, no.~1, p.~108613, 2024.

\bibitem{chen2024unmasking}
F.~Chen, L.~Wang, J.~Hong, J.~Jiang, and L.~Zhou, ``Unmasking bias in artificial intelligence: a systematic review of bias detection and mitigation strategies in electronic health record-based models,'' {\em Journal of the American Medical Informatics Association}, vol.~31, no.~5, pp.~1172--1183, 2024.

\bibitem{hung2023developing}
S.-K. Hung, C.-C. Wu, A.~Singh, J.-H. Li, C.~Lee, E.~H. Chou, A.~Pekosz, R.~Rothman, and K.-F. Chen, ``Developing and validating clinical features-based machine learning algorithms to predict influenza infection in influenza-like illness patients,'' {\em biomedical journal}, vol.~46, no.~5, p.~100561, 2023.

\bibitem{tso2022machine}
C.~F. Tso, C.~Lam, J.~Calvert, and Q.~Mao, ``Machine learning early prediction of respiratory syncytial virus in pediatric hospitalized patients,'' {\em Frontiers in Pediatrics}, vol.~10, p.~886212, 2022.

\bibitem{zhang2020machine}
L.~Zhang, Y.~Wang, M.~Niu, C.~Wang, and Z.~Wang, ``Machine learning for characterizing risk of type 2 diabetes mellitus in a rural chinese population: The henan rural cohort study,'' {\em Scientific reports}, vol.~10, no.~1, p.~4406, 2020.

\bibitem{wang2021characteristics}
X.~Wang, Y.~Xiao, X.~Xu, L.~Guo, Y.~Yu, N.~Li, and C.~Xu, ``Characteristics of fecal microbiota and machine learning strategy for fecal invasive biomarkers in pediatric inflammatory bowel disease,'' {\em Frontiers in Cellular and Infection Microbiology}, vol.~11, p.~711884, 2021.

\bibitem{thomas2019metagenomic}
A.~M. Thomas, P.~Manghi, F.~Asnicar, E.~Pasolli, F.~Armanini, M.~Zolfo, F.~Beghini, S.~Manara, N.~Karcher, C.~Pozzi, {\em et~al.}, ``Metagenomic analysis of colorectal cancer datasets identifies cross-cohort microbial diagnostic signatures and a link with choline degradation,'' {\em Nature medicine}, vol.~25, no.~4, pp.~667--678, 2019.

\bibitem{lee2023machine}
Y.~Lee, M.~Cappellato, and B.~Di~Camillo, ``Machine learning--based feature selection to search stable microbial biomarkers: application to inflammatory bowel disease,'' {\em GigaScience}, vol.~12, p.~giad083, 2023.

\bibitem{aryal2020machine}
S.~Aryal, A.~Alimadadi, I.~Manandhar, B.~Joe, and X.~Cheng, ``Machine learning strategy for gut microbiome-based diagnostic screening of cardiovascular disease,'' {\em Hypertension}, vol.~76, no.~5, pp.~1555--1562, 2020.

\bibitem{lundberg2017unified}
S.~M. Lundberg and S.-I. Lee, ``A unified approach to interpreting model predictions,'' {\em Advances in neural information processing systems}, vol.~30, 2017.

\bibitem{aghababa2024exploration}
M.~P. Aghababa and J.~Andrysek, ``Exploration and demonstration of explainable machine learning models in prosthetic rehabilitation-based gait analysis,'' {\em Plos one}, vol.~19, no.~4, p.~e0300447, 2024.

\bibitem{fuse2024development}
Y.~Fuse, K.~Ishii, F.~Kanamori, S.~Oyama, T.~Imaizumi, Y.~Araki, K.~Yokoyama, S.~Takasu, Y.~Seki, and R.~Saito, ``Development and validation of machine learning models to predict postoperative infarction in moyamoya disease,'' {\em Journal of Neurosurgery}, vol.~1, no.~aop, pp.~1--9, 2024.

\bibitem{you2024development}
C.~You, J.~Ren, L.~Cheng, C.~Peng, P.~Lu, K.~Guo, F.~Zhong, J.~Wang, X.~Gao, J.~Cao, {\em et~al.}, ``Development and validation of a machine learning-based postoperative prognostic model for plasma cell neoplasia with spinal lesions as initial clinical manifestations: a single-center cohort study,'' {\em European Spine Journal}, pp.~1--11, 2024.

\bibitem{fan2024predicting}
J.~Fan, S.~Shi, H.~Xiang, L.~Fu, Y.~Duan, D.~Cao, and H.~Lu, ``Predicting elimination of small-molecule drug half-life in pharmacokinetics using ensemble and consensus machine learning methods,'' {\em Journal of Chemical Information and Modeling}, 2024.

\bibitem{he2024ultrasound}
Y.~He, B.~Zheng, W.~Peng, Y.~Chen, L.~Yu, W.~Huang, and G.~Qin, ``An ultrasound-based ensemble machine learning model for the preoperative classification of pleomorphic adenoma and warthin tumor in the parotid gland,'' {\em European Radiology}, pp.~1--15, 2024.

\bibitem{liu2024development}
Y.-Q. Liu, W.-H. Yuan, Y.~Tao, L.~Zhao, and W.-L. Guo, ``Development of a machine learning model and nomogram to predict seizures in children with covid-19: a two-center study,'' {\em Journal of Tropical Pediatrics}, vol.~70, no.~3, p.~fmae011, 2024.

\bibitem{li2024insights}
J.~Li, W.~Qin, B.~Zhu, T.~Ruan, Z.~Hua, H.~Du, S.~Dong, and J.~Fang, ``Insights into the transformation of natural organic matter during uv/peroxydisulfate treatment by ft-icr ms and machine learning: Non-negligible formation of organosulfates,'' {\em Water Research}, vol.~256, p.~121564, 2024.

\bibitem{venturini2024predicting}
M.~Venturini, I.~Van~Keilegom, W.~De~Corte, and C.~Vens, ``Predicting time-to-intubation after critical care admission using machine learning and cured fraction information,'' {\em Artificial Intelligence in Medicine}, p.~102817, 2024.

\bibitem{yu2024exploring}
Y.~Yu, Y.~Xia, and G.~Liang, ``Exploring novel lead scaffolds for sglt2 inhibitors: Insights from machine learning and molecular dynamics simulations,'' {\em International Journal of Biological Macromolecules}, p.~130375, 2024.

\bibitem{abujaber2024machine}
A.~A. Abujaber, I.~Albalkhi, Y.~Imam, A.~Nashwan, N.~Akhtar, and I.~M. Alkhawaldeh, ``Machine learning-based prognostication of mortality in stroke patients,'' {\em Heliyon}, vol.~10, no.~7, 2024.

\bibitem{vimbi2024interpreting}
V.~Vimbi, N.~Shaffi, and M.~Mahmud, ``Interpreting artificial intelligence models: a systematic review on the application of lime and shap in alzheimer’s disease detection,'' {\em Brain Informatics}, vol.~11, no.~1, p.~10, 2024.

\bibitem{wang2024quantifying}
L.~Wang, H.~Wang, F.~D’Angelo, L.~Curtin, C.~P. Sereduk, G.~D. Leon, K.~W. Singleton, J.~Urcuyo, A.~Hawkins-Daarud, P.~R. Jackson, {\em et~al.}, ``Quantifying intra-tumoral genetic heterogeneity of glioblastoma toward precision medicine using mri and a data-inclusive machine learning algorithm,'' {\em Plos one}, vol.~19, no.~4, p.~e0299267, 2024.

\bibitem{shinohara2024re}
I.~Shinohara, Y.~Mifune, A.~Inui, H.~Nishimoto, T.~Yoshikawa, T.~Kato, T.~Furukawa, S.~Tanaka, M.~Kusunose, Y.~Hoshino, {\em et~al.}, ``Re-tear after arthroscopic rotator cuff tear surgery: risk analysis using machine learning,'' {\em Journal of Shoulder and Elbow Surgery}, vol.~33, no.~4, pp.~815--822, 2024.

\bibitem{huang2024new}
F.~Huang and X.~Zhang, ``A new interpretable streamflow prediction approach based on swat-bilstm and shap,'' {\em Environmental Science and Pollution Research}, pp.~1--13, 2024.

\bibitem{ma2024integrated}
K.~Ma, ``Integrated hybrid modeling and shap (shapley additive explanations) to predict and explain the adsorption properties of thermoplastic polyurethane (tpu) porous materials,'' {\em RSC advances}, vol.~14, no.~15, pp.~10348--10357, 2024.

\bibitem{rodriguez2024machine}
P.~Rodr{\'\i}guez-Belenguer, J.~L. Pi{\~n}ana, M.~S{\'a}nchez-Monta{\~n}{\'e}s, E.~Soria-Olivas, M.~Mart{\'\i}nez-Sober, and A.~J. Serrano-L{\'o}pez, ``A machine learning approach to identify groups of patients with hematological malignant disorders,'' {\em Computer Methods and Programs in Biomedicine}, vol.~246, p.~108011, 2024.

\bibitem{hernandez2024explainable}
M.~C. Hernandez, C.~Chen, A.~Nguyen, K.~Choong, C.~Carlin, R.~A. Nelson, L.~A. Rossi, N.~Seth, K.~McNeese, B.~Yuh, {\em et~al.}, ``Explainable machine learning model to preoperatively predict postoperative complications in inpatients with cancer undergoing major operations,'' {\em JCO Clinical Cancer Informatics}, vol.~8, p.~e2300247, 2024.

\bibitem{tong2024machine}
C.~Tong, X.~Du, Y.~Chen, K.~Zhang, M.~Shan, Z.~Shen, H.~Zhang, and J.~Zheng, ``Machine learning prediction model of major adverse outcomes after pediatric congenital heart surgery: a retrospective cohort study,'' {\em International Journal of Surgery}, vol.~110, no.~4, pp.~2207--2216, 2024.

\bibitem{guo2024application}
Q.-H. Guo, F.-C. Xie, F.-M. Zhong, W.~Wen, X.-R. Zhang, X.-J. Yu, X.-L. Wang, B.~Huang, L.-P. Li, and X.-Z. Wang, ``Application of interpretable machine learning algorithms to predict distant metastasis in ovarian clear cell carcinoma,'' {\em Cancer Medicine}, vol.~13, no.~7, p.~e7161, 2024.

\bibitem{wang2024exploring}
S.~Wang, T.~Zhang, Z.~Li, and J.~Hong, ``Exploring pollutant joint effects in disease through interpretable machine learning,'' {\em Journal of Hazardous Materials}, p.~133707, 2024.

\bibitem{meng2024application}
L.~Meng, P.~Zhu, and K.~Xia, ``Application value of the automated machine learning model based on modified ct index combined with serological indices in the early prediction of lung cancer,'' {\em Frontiers in Public Health}, vol.~12, p.~1368217, 2024.

\bibitem{li2024incremental}
S.~Li, Z.~Zhou, M.~Gao, Z.~Liao, K.~He, W.~Qu, J.~Li, I.~R. Kamel, Q.~Chu, Q.~Zhang, {\em et~al.}, ``Incremental value of automatically segmented perirenal adipose tissue for pathological grading of clear cell renal cell carcinoma: a multicenter cohort study,'' {\em International Journal of Surgery}, pp.~10--1097, 2024.

\bibitem{nagy2024predicting}
M.~Nagy, A.~M. Onder, D.~Rosen, C.~Mullett, A.~Morca, and O.~Baloglu, ``Predicting pediatric cardiac surgery-associated acute kidney injury using machine learning,'' {\em Pediatric Nephrology}, vol.~39, no.~4, pp.~1263--1270, 2024.

\bibitem{rattsev2024incorporation}
I.~Rattsev, V.~Stearns, A.~L. Blackford, D.~L. Hertz, K.~L. Smith, J.~M. Rae, and C.~O. Taylor, ``Incorporation of emergent symptoms and genetic covariates improves prediction of aromatase inhibitor therapy discontinuation,'' {\em JAMIA open}, vol.~7, no.~1, p.~ooae006, 2024.

\bibitem{chen2024machine}
C.~Chen, Y.~Wang, X.~Yang, M.~Zhang, J.~He, L.~Yang, L.~Qin, B.~Chen, B.~Chen, and Q.~Wang, ``Machine learning-based prediction of intraoperative red blood cell transfusion in aortic valve replacement surgery.,'' {\em Clinical Laboratory}, vol.~70, no.~4, 2024.

\bibitem{lee2024essential}
Y.-H. Lee, J.~Chang, J.-E. Lee, Y.-S. Jung, D.~Lee, and H.-S. Lee, ``Essential elements of physical fitness analysis in male adolescent athletes using machine learning,'' {\em Plos one}, vol.~19, no.~4, p.~e0298870, 2024.

\bibitem{cheng2024early}
Y.-W. Cheng, P.-C. Kuo, S.-H. Chen, Y.-T. Kuo, T.-L. Liu, W.-S. Chan, K.-C. Chan, and Y.-C. Yeh, ``Early prediction of mortality at sepsis diagnosis time in critically ill patients by using interpretable machine learning,'' {\em Journal of Clinical Monitoring and Computing}, vol.~38, no.~2, pp.~271--279, 2024.

\bibitem{chahine2024machine}
Y.~Chahine, T.~Afroze, S.~F. Bifulco, D.~V. Tekmenzhi, M.~Jafarvand, P.~M. Boyle, and N.~Akoum, ``Machine learning identifies esophageal luminal temperature patterns associated with thermal injury in catheter ablation for atrial fibrillation,'' {\em Journal of Cardiovascular Electrophysiology}, 2024.

\bibitem{xu2024interactive}
X.~Xu, W.~Gu, X.~Shen, Y.~Liu, S.~Zhai, C.~Xu, G.~Cui, and L.~Xiao, ``An interactive web application to identify early parkinsonian non-tremor-dominant subtypes,'' {\em Journal of Neurology}, pp.~1--9, 2024.

\bibitem{pandey2024functional}
A.~K. Pandey, S.~C. Nayak, and S.-H. Kim, ``Functional link hybrid artificial neural network for predicting continuous biohydrogen production in dynamic membrane bioreactor,'' {\em Bioresource Technology}, p.~130496, 2024.

\bibitem{gholi2024explainable}
F.~Gholi Zadeh~Kharrat, C.~Gagne, A.~Lesage, G.~Gari{\'e}py, J.-F. Pelletier, C.~Brousseau-Paradis, L.~Rochette, E.~Pelletier, P.~L{\'e}vesque, M.~Mohammed, {\em et~al.}, ``Explainable artificial intelligence models for predicting risk of suicide using health administrative data in quebec,'' {\em PLoS one}, vol.~19, no.~4, p.~e0301117, 2024.

\bibitem{ciciora2024social}
D.~Ciciora, E.~V{\'a}squez, E.~Valachovic, L.~Hou, Y.~Zheng, H.~Xu, X.~Jiang, K.~Huang, K.~P. Gabriel, H.-W. Deng, {\em et~al.}, ``Social and behavior factors of alzheimer's disease and related dementias: A national study in the us,'' {\em American Journal of Preventive Medicine}, vol.~66, no.~4, pp.~573--581, 2024.

\bibitem{mallick2024game}
J.~Mallick, M.~Alkahtani, H.~T. Hang, and C.~K. Singh, ``Game-theoretic optimization of landslide susceptibility mapping: a comparative study between bayesian-optimized basic neural network and new generation neural network models,'' {\em Environmental Science and Pollution Research}, vol.~31, no.~20, pp.~29811--29835, 2024.

\bibitem{cao2024explainable}
L.~Cao, X.~Ma, W.~Huang, G.~Xu, Y.~Wang, M.~Liu, S.~Sheng, and K.~Mao, ``An explainable artificial intelligence model to predict malignant cerebral edema after acute anterior circulating large-hemisphere infarction,'' {\em European Neurology}, vol.~87, no.~2, pp.~54--66, 2024.

\bibitem{liu2024improving}
X.~Liu, H.~Niu, and J.~Peng, ``Improving predictions: Enhancing in-hospital mortality forecast for icu patients with sepsis-induced coagulopathy using a stacking ensemble model,'' {\em Medicine}, vol.~103, no.~14, p.~e37634, 2024.

\bibitem{lun2024prediction}
Y.~Lun, H.~Yuan, P.~Ma, J.~Chen, P.~Lu, W.~Wang, R.~Liang, J.~Zhang, W.~Gao, X.~Ding, {\em et~al.}, ``A prediction model based on random survival forest analysis of the overall survival of elderly female papillary thyroid carcinoma patients: a seer-based study,'' {\em Endocrine}, pp.~1--9, 2024.

\bibitem{rajwa2024identification}
B.~Rajwa, M.~M.~A. Naved, M.~Adibuzzaman, A.~Y. Grama, B.~A. Khan, M.~M. Dundar, and J.-C. Rochet, ``Identification of predictive patient characteristics for assessing the probability of covid-19 in-hospital mortality,'' {\em PLOS Digital Health}, vol.~3, no.~4, p.~e0000327, 2024.

\bibitem{ekeany2020borutashap}
E.~Ekeany, ``Boruta-shap: A tree based feature selection tool which combines both the boruta feature selection algorithm with shapley values,'' 2020.
\newblock Accessed: 2024-01-02.

\bibitem{calzolari2020shapicant}
M.~Calzolari, ``manuel-calzolari/shapicant: Feature selection package based on shap and target permutation, for pandas and spark,'' 2020.
\newblock Accessed: 2024-01-02.

\bibitem{verhaeghe2022powershap}
J.~Verhaeghe, J.~Van Der~Donckt, F.~Ongenae, and S.~Van~Hoecke, ``Powershap: A power-full shapley feature selection method,'' in {\em Joint European conference on machine learning and knowledge discovery in databases}, pp.~71--87, Springer, 2022.

\bibitem{madakkatel2024llpowershap}
I.~Madakkatel and E.~Hypp{\"o}nen, ``Llpowershap: logistic loss-based automated shapley values feature selection method,'' {\em BMC Medical Research Methodology}, vol.~24, no.~1, p.~247, 2024.

\bibitem{kraev2410shap}
E.~Kraev, B.~Koseoglu, L.~Traverso, and M.~Topiwalla, ``Shap-select: Lightweight feature selection using shap values and regression. arxiv 2024,'' {\em arXiv preprint arXiv:2410.06815}.

\bibitem{teneggi2022shap}
J.~Teneggi, B.~Bharti, Y.~Romano, and J.~Sulam, ``Shap-xrt: The shapley value meets conditional independence testing,'' {\em arXiv preprint arXiv:2207.07038}, 2022.

\bibitem{lee2024validity}
Y.~Lee, G.~Baruzzo, J.~Kim, J.~Seo, and B.~Di~Camillo, ``Validity of feature importance in low-performing machine learning for tabular biomedical data,'' {\em arXiv preprint arXiv:2409.13342}, 2024.

\bibitem{yuan2022empirical}
H.~Yuan, M.~Liu, L.~Kang, C.~Miao, and Y.~Wu, ``An empirical study of the effect of background data size on the stability of shapley additive explanations (shap) for deep learning models,'' {\em arXiv preprint arXiv:2204.11351}, 2022.

\bibitem{cifuentes2024distributional}
S.~Cifuentes, L.~Bertossi, N.~Pardal, S.~Abriola, M.~V. Martinez, and M.~Romero, ``The distributional uncertainty of the shap score in explainable machine learning,'' {\em arXiv preprint arXiv:2401.12731}, 2024.

\bibitem{zeng2024enhancing}
X.~Zeng, ``Enhancing the interpretability of shap values using large language models,'' {\em arXiv preprint arXiv:2409.00079}, 2024.

\bibitem{hsu2024decoding}
C.-C. Hsu, I.-Z. Wu, and S.-M. Liu, ``Decoding ai complexity: Shap textual explanations via llm for improved model transparency,'' in {\em 2024 International Conference on Consumer Electronics-Taiwan (ICCE-Taiwan)}, pp.~197--198, IEEE, 2024.

\bibitem{albert2024_metabolic}
A.~ANTONY, ``Metabolic syndrome,'' 2023.

\bibitem{patel2015reliable}
B.~C. Patel and G.~Sinha, ``Reliable computer-aided diagnosis system using region based segmentation of mammographic breast cancer images,'' {\em Mathematical Methods and Systems in Science and Engineering}, vol.~296, 2015.

\bibitem{https://doi.org/10.24432/c5z89r}
G.~J. D.~Chicco, ``Heart failure clinical records,'' 2020.

\bibitem{antal2014diabetic}
B.~Antal and A.~Hajdu, ``Diabetic retinopathy debrecen data set,'' {\em UCI Mach. Learn. Repos}, 2014.

\bibitem{lee2023suggestion}
Y.~Lee and J.~Seo, ``Suggestion of statistical validation on feature importance of machine learning,'' in {\em 2023 45th Annual International Conference of the IEEE Engineering in Medicine \& Biology Society (EMBC)}, pp.~1--4, IEEE, 2023.

\bibitem{lundberg2019explainable}
S.~M. Lundberg, G.~Erion, H.~Chen, A.~DeGrave, J.~M. Prutkin, B.~Nair, R.~Katz, J.~Himmelfarb, N.~Bansal, and S.-I. Lee, ``Explainable ai for trees: From local explanations to global understanding,'' {\em arXiv preprint arXiv:1905.04610}, 2019.

\bibitem{lee4744730biomarker}
Y.~Lee, J.~Seo, and B.~Di~Camillo, ``Biomarker selection using shap based binarization: Application to microbiota-ibd data,'' {\em Available at SSRN 4744730}.

\bibitem{slack2020fooling}
D.~Slack, S.~Hilgard, E.~Jia, S.~Singh, and H.~Lakkaraju, ``Fooling lime and shap: Adversarial attacks on post hoc explanation methods,'' in {\em Proceedings of the AAAI/ACM Conference on AI, Ethics, and Society}, pp.~180--186, 2020.

\bibitem{adebayo2018sanity}
J.~Adebayo, J.~Gilmer, M.~Muelly, I.~Goodfellow, M.~Hardt, and B.~Kim, ``Sanity checks for saliency maps,'' {\em Advances in neural information processing systems}, vol.~31, 2018.

\end{thebibliography}

\newpage

\section{Biography Section}

\begin{IEEEbiography}
[{\includegraphics[width=1in,height=1.25in,clip,keepaspectratio]{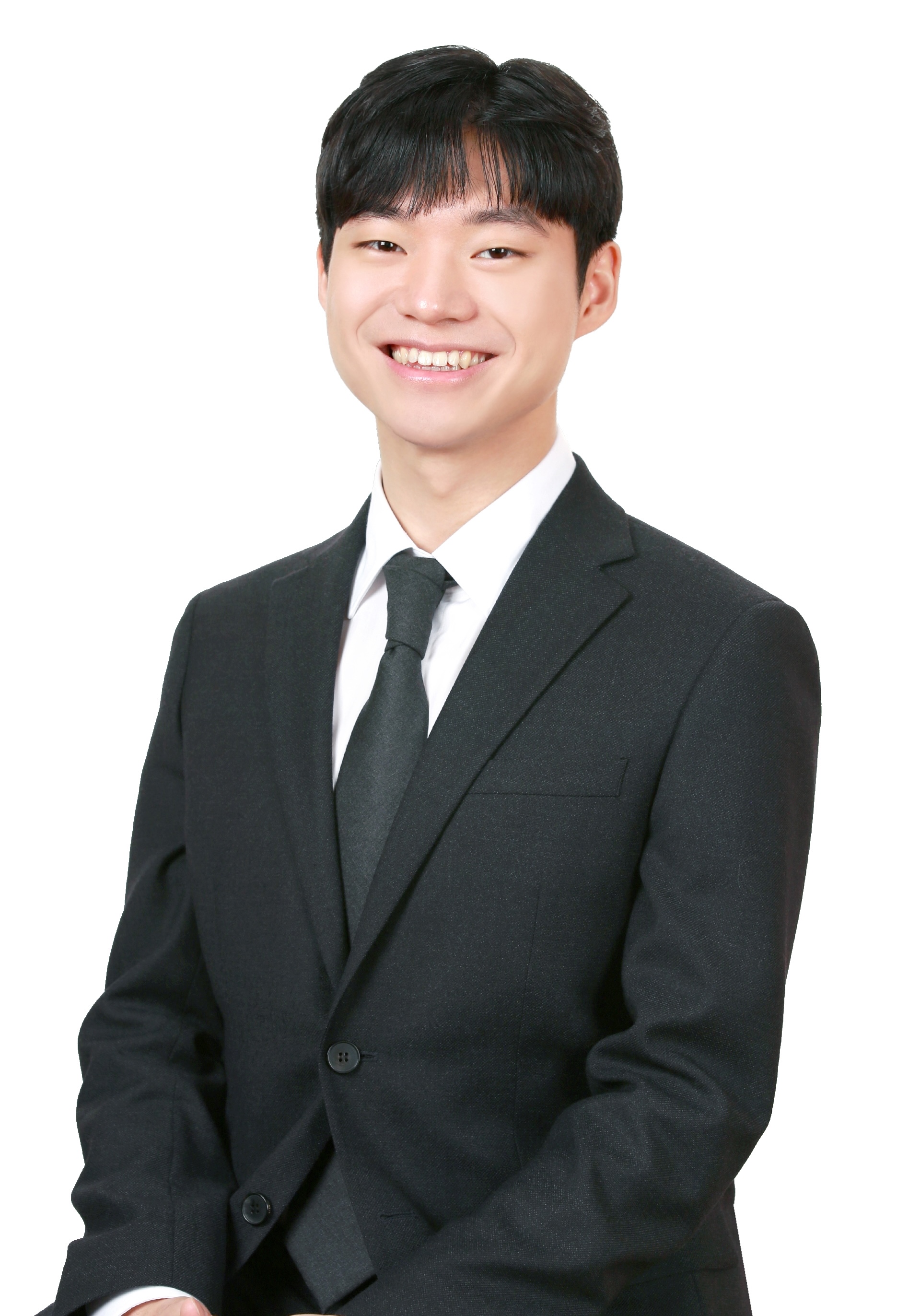}}] {Kyungjin Kim} received the B.S. degree from Seoul National University, where he is currently pursuing a Ph.D. in Electrical and Computer Engineering. He is currently a visiting researcher at the Human-Computer Interaction Institute (HCII), Carnegie Mellon University. His research focuses on enhancing the robustness and interpretability of Clinical Decision Support Systems (CDSS) by integrating Explainable AI (XAI) and interactive data visualization. At CMU, he explores human-AI interaction frameworks to facilitate the intuitive interpretation of complex machine learning models within data-driven environments.
\end{IEEEbiography}

\begin{IEEEbiography}[{\includegraphics[width=1in,height=1.25in,clip,keepaspectratio]{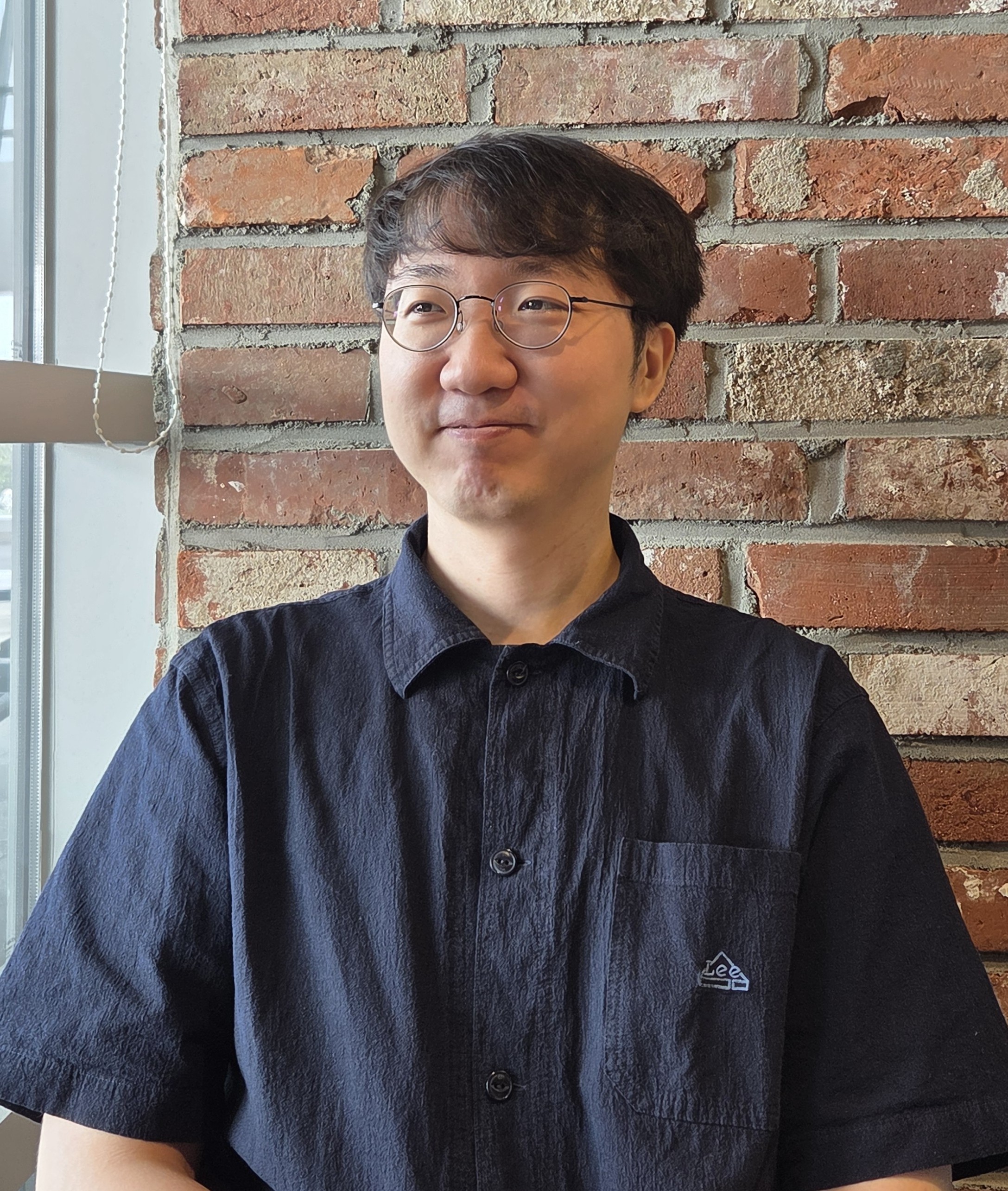}}]{Youngro Lee} received his Ph.D. in Electrical and Computer Engineering from Seoul National University in February 2025, where he currently continues as a postdoctoral researcher. His research centers on applying machine learning and explainable AI to diverse medical and biological domains, including microbiome biomarker discovery, influenza-like illness decision support systems, and knowledge discovery for metabolic health. He has collaborated closely with clinicians across multiple institutions to ensure the validity and interpretability of AI models. His broader interest lies in developing trustworthy and reproducible machine learning applications in healthcare and computational biology. He also conducted a research visit at the University of Padova in 2021 and has since continued collaborative research on computational biology and machine learning pipelines.
\end{IEEEbiography}

\begin{IEEEbiography}[{\includegraphics[width=1in,height=1.25in,clip,keepaspectratio]{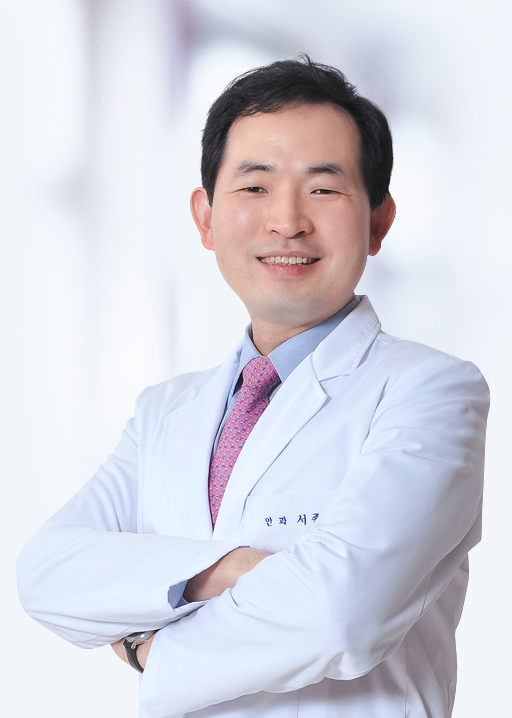}}]{Jongmo Seo} (Member, IEEE) received the M.D. degree in medicine in 1996 and the M.S. and Ph.D. degrees in biomedical engineering in 2002 and 2005, respectively, all from Seoul National University, Seoul, South Korea. He is a board‑certified ophthalmologist and is currently with the Department of Electrical and Computer Engineering, Seoul National University, where he has been a faculty member since 2008. He is also a professor in the Interdisciplinary Program in Medical Informatics at Seoul National University College of Medicine, where he serves as the head of the program. He is a member of the IEEE Engineering in Medicine and Biology Society (EMBS). His research interests include neural interfaces, artificial vision, bioelectronics, and the development of implantable medical devices such as artificial retina.

\end{IEEEbiography}

\vfill

\end{document}